\documentclass[sigconf]{acmart}

\AtBeginDocument{%
  \providecommand\BibTeX{{%
    \normalfont B\kern-0.5em{\scshape i\kern-0.25em b}\kern-0.8em\TeX}}}

\setcopyright{acmcopyright}
\copyrightyear{2020}
\acmYear{2020}
\setcopyright{acmcopyright}
\acmConference[MM '20]{Proceedings of the 28th ACM International Conference on Multimedia}{October 12--16, 2020}{Seattle, WA, USA}
\acmBooktitle{Proceedings of the 28th ACM International Conference on Multimedia (MM '20), October 12--16, 2020, Seattle, WA, USA}
\acmPrice{15.00}
\acmDOI{10.1145/3394171.3413881}
\acmISBN{978-1-4503-7988-5/20/10}

\usepackage[pdftex]{graphicx}
\usepackage{subcaption}

\begin{document}
\fancyhead{}

\title{Building Movie Map - A Tool for Exploring Areas in a City - \\ and its Evaluations}

\author{Naoki Sugimoto}
\email{sugimoto@hal.t.u-tokyo.ac.jp}
\affiliation{%
  \institution{Department of Information and Communication Engineering, The University of Tokyo}
  \city{Tokyo}
  \country{Japan}
  \postcode{113-8656}
}

\author{Yoshihito Ebine}
\email{yoshi.ebine@vteclaboratories.co.jp}
\affiliation{%
  \institution{VTEC Laboratories Inc.}
  \city{Tokyo}
  \country{Japan}
  \postcode{102-0083}
}

\author{Kiyoharu Aizawa}
\email{aizawa@hal.t.u-tokyo.ac.jp}
\affiliation{%
  \institution{Department of Information and Communication Engineering, The University of Tokyo}
  \city{Tokyo}
  \country{Japan}
  \postcode{113-8656}
}

\renewcommand{\shortauthors}{Sugimoto et al.}

\begin{abstract}
We propose a new Movie Map, which will enable users to explore a given city area using omnidirectional videos.
Only one Movie Map prototype was developed in the 1980s; it was developed with analog video technology. 
Later, Google Street View (GSV) provided interactive panoramas from positions along streets around the world in Google Maps.
Despite the wide use of GSV, it provides sparse images of streets, which often confuses users and lowers user satisfaction.
Movie Map’s use of videos instead of sparse images dramatically improves the user experience.
Thus, we improve the Movie Map using state-of-the-art technology.
We propose a new Movie Map system, with an interface for exploring cities.
The system consists of four stages; acquisition, analysis, management, and interaction.
In the acquisition stage, omnidirectional videos are taken along streets in target areas.
Frames of the video are localized on the map, 
intersections are detected, and videos are segmented.
Turning views at intersections are subsequently generated.
By connecting the video segments following the specified movement in an area, we can view the streets better.
The interface allows for easy exploration of a target area, and it can show virtual billboards of stores in the view.
We conducted user studies to compare our system to the GSV in a scenario where users could freely move and explore to find a landmark.
The experiment showed that our system had a better user experience than GSV.
\end{abstract}




\begin{teaserfigure}
\begin{center}
  \includegraphics[keepaspectratio,scale=0.3]{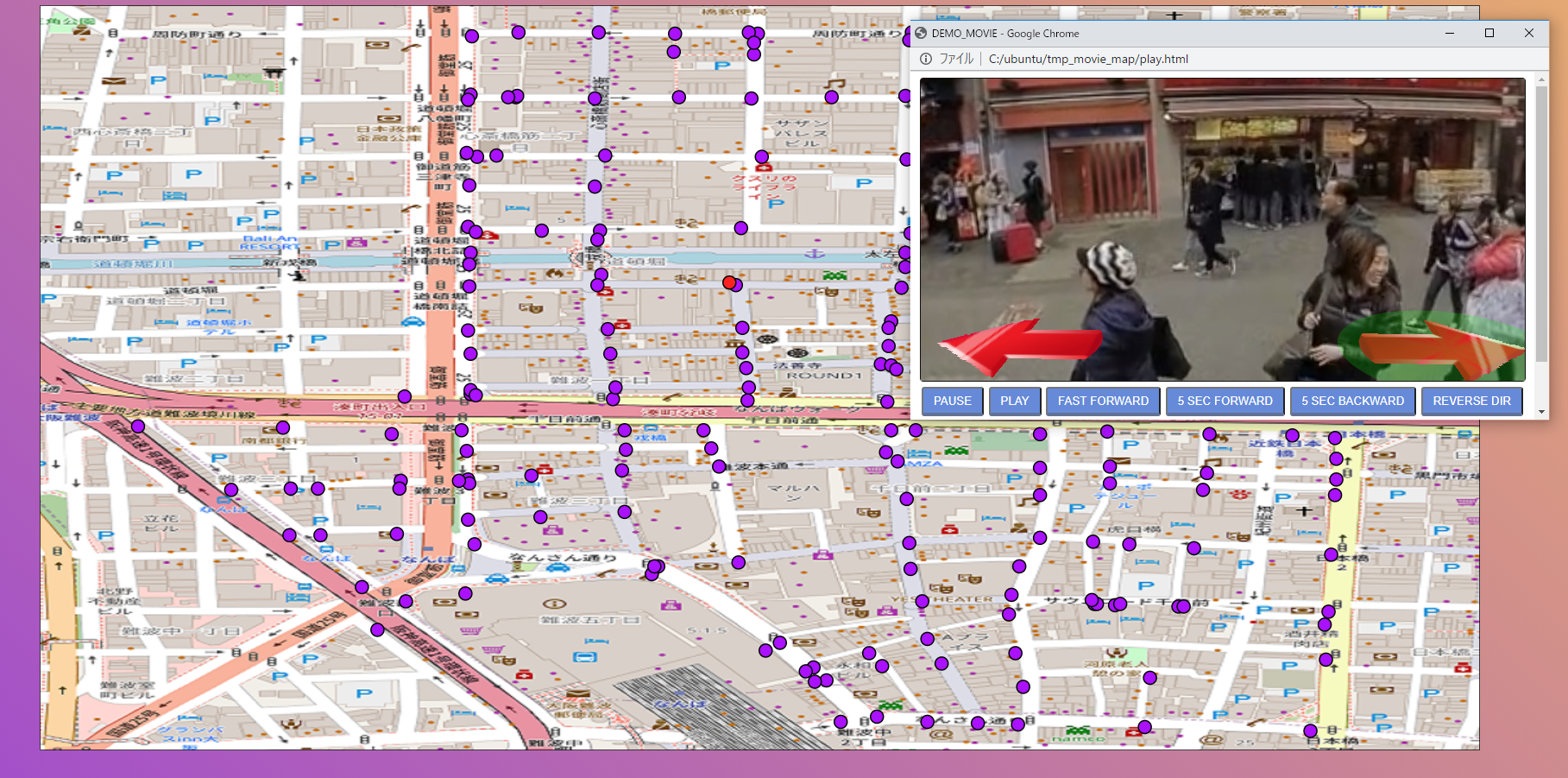}
  \caption{An exploring interface of the system for the areas around Namba}
\label{interface}
\end{center}
\end{teaserfigure}

\maketitle

\section{Introduction}

There is a significant demand for obtaining information on areas that are yet to be visited.
This can be achieved by accessing a map of the area, reading documents regarding regional features or tourist information, and viewing captured pictures and movies of that area. 

Each of these methods has advantages and disadvantages.
A map can provide a bird's-eye view of the entire target region at a single glance; however, this information is limited, depending on the type of map~\cite{map_advantage}.
Documents provide detailed descriptions of areas.
Nevertheless, it is generally necessary to refer to several documents to grasp the overall image of the area.
Contrastingly, pictures and movies are excellent for elucidating the view and atmosphere of a location.
Nonetheless, it is not easy to grasp the entire view of an area using this approach.

Currently, there are various interfaces of location information that combine different types of data of an area~\cite{yahoo} ~\cite{geographical_survey} ~\cite{bing}\\~\cite{mapillary} ~\cite{mapfan} ~\cite{offmaps}~\cite{sugimoto_icmr}.
For example, on websites such as the Japanese Geographical Survey Institute \cite{geographical_survey}, information indicating the characteristics of an area can be visualized via graphs, etc., by specifying a prefecture on the map. This is realized via an interface that displays detailed information of the target areas, while grasping a huge geographic overview of Japan.

Google Street View (GSV)~\cite{street_view} is an interface that combines maps and images.
It provides users with omnidirectional images corresponding to a chosen location on the map.
Users interact with these images, and movements on the map are updated on the displayed content of those images.
This creates a realistic user experience as it enables users to experience the place as if they were there.
GSV is widely used and offers convenience for tasks such as guidance and pre-learning about a target area~\cite{use_gsv}.

However, GSV is not a perfect interface for the abovementioned tasks.
In GSV, street images are significantly sparse. 
To view all the images along a route on GSV, users need to several transitions between those images; consequently, the user has to interact multiple times with the system~\cite{gsv_dis}.
This is quite tedious and occasionally difficult.
Such image transitions may result in the user getting lost or being led along an incorrect direction.
Additionally, because of the interval between images, the user does not experience a continuous movement.

However, the use of videos instead of sparse images can solve these problems that are experienced by GSV users.
When using videos, we can select the starting point and direction on the map, and play back a video along the streets of the map along the desired direction.
It is possible to place video transitions at the intersections and subsequently change the direction of movement through interactions.
Hence, the use of videos can eliminate the need for excessive interactions and also enable users to experience continuous movement.

Forty years ago, a research project prototyped Movie Map\cite{movie_map} based on analog video technology was constructed.
Movie Map was used to play street movies corresponding to the directions of a car in Aspen city.
These street segment movies were recorded on multiple optical disks and were played according to the user's inputs.
The Movie Map was produced only once.
This was because it was not easy to simultaneously acquire video and geotags.
Furthermore, multimedia interaction machines were not readily available.
Therefore, GSV has been more successful in recent years. 

We build our Movie Map by incorporating current technology.
This renovated version can be used for pre-learning directions, or virtual tours for walkers in certain areas, such as commercial areas around a station.
We acquire omnidirectional videos along the streets of a target area, analyze the camera positions of videos using the visual simultaneous localization and mapping (vSLAM) technique, and associate the corresponding video frames with locations on the map.
We detect intersection frames among videos based on location information and visual features, and the segmented videos are subsequently organized by intersections.
Thereafter, we build an interface to intuitively explore a target area.
Thus, the proposed Movie Map enables users to easily move along the streets in an area by synthesizing views connected via video segments. 

We evaluate the effectiveness of route synthesis by connecting video segments at the intersections and using the rotating synthesized views, obtained by blending intersection frames, of the two relevant videos.
We also evaluate the advantages and disadvantages as well as the user satisfaction of the proposed interface for exploring tasks and compare them to GSV.
 
Our contributions are as follows.
\begin{itemize}
\item We build a Movie Map that allows the user to interactively explore a certain target area displaying omnidirectional street videos captured along the streets in the target area.
After capturing the videos in the area, the processing required for the video database is automated except for the assigning of two reference points per video.
We capture the omnidirectional videos of two directions along streets in areas around Kyoto station and Namba station.
The system is easily applicable to different areas.

\item We produce a natural transition from streets to other streets at intersections by generating turning views.
In the user study, turning views were highly evaluated compared to directly switching and simple rotation.

\item We evaluated our system against GSV under a scenario where users explore an area by looking for a specific location; our system was evaluated higher in terms of exploring comfort.

\item As an extension, we use virtual billboards in the view of the Movie Map.
The billboards are shown at the locations of the shops and stores in the views.
When the billboards are clicked, their associated information pops up.
\end{itemize}

Recent work by Sugimoto et al.~\cite{sugimoto_icmr} demonstrated a Movie Map that can show a synthesis route video by determining the route on a map in advance.
This does not allow users to freely explore the specific area.
In this work, the proposed system enables users to explore the area without specifying the route in advance.

\section{Related Work}

\subsection{Movie Map}

The concept of a Movie Map was originally proposed by Lippmann four decades ago~\cite{movie_map} and the prototype system is well known as the Aspen Movie Map for the Aspen city area.
It is an interactive map, built by using video disc technology to engage the user in a simulated "drive" through an unfamiliar space. 
In this system, panoramic images were captured by four cameras that were placed at $90^\circ$ intervals in a horizontal circle every 10 feet; their precise locations were measured via GPS.
Images along roadways and at intersections were stored in several optical disk drives. A user could interactively explore the area on a touch screen by specifying the direction of movement.
The Movie Map provided a methodology for visualizing a certain area by only using multiple video segments. 
However, the technology used at that time was not sufficiently advanced and therefore the map could not be scaled.
Until now, minimal research in this field has been conducted; they are summarized in~\cite{movie_map_naimark}.
There are limited examples of Movie Maps~\cite{see_banff}~\cite{sugimoto_icmr}.
~\cite{see_banff} replay captured route movies without any route synthesis.
Thus, it cannot be efficiently generalized for a Movie Map that covers a certain area.
Sugimoto et al.~\cite{sugimoto_icmr} present a Movie Map with a interface in which a user specifies his/her route on a map in advance.
Consequently, the user cannot freely explore the target-area.

GSV emerged in 2007; it was initially used on several cities in the US before expanding globally~\cite{gsv_area}.
It provides an interactive slide-show type view of maps.
Presently, unlike GSV, the use of a Movie Map is highly uncommon.

GSV is not necessarily better than Movie Map regarding user experience.
\cite{movie_map} claimed that Movie Map allows the user to experience an area as if they were driving through the area.
Such an experience is provided to the user by using continuous visual information; this cannot be obtained by GSV.
In this regard, Movie Maps are still worth studying.
Therefore, we take advantage of today's technology to improve a Movie Map for walkers in certain city areas.

\subsection{Visual Simultaneous Localization and Mapping}

\begin{figure}[t]
\begin{center}
  \includegraphics[keepaspectratio,scale=0.35]{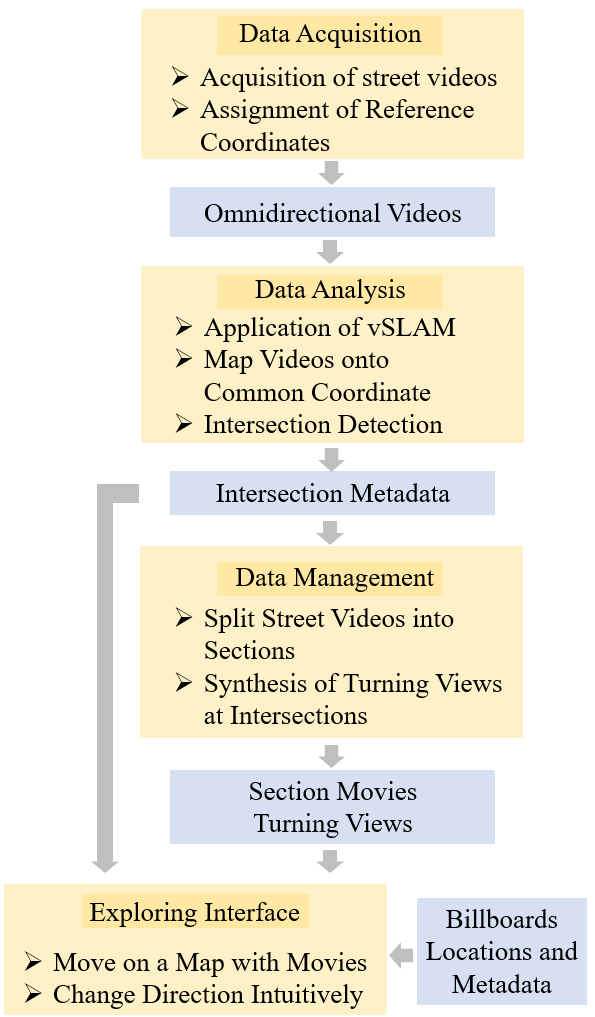}
  \caption{Outline of the system}
\label{system_abstract}
\end{center}
\end{figure}

Visual simultaneous localization and mapping (vSLAM) is a technique used for 3D reconstruction and camera position estimation by using a video captured by a single camera~\cite{slam1}~\cite{slam2}~\cite{slam3}.
Visual features and camera parameters are used for the optimization of relative changes in the camera position and orientation between consecutive frames.

In our study, we apply OpenVSLAM~\cite{openvslam2019}, which is one of the opensource vSLAM software, to capture street videos and to estimate their relative trajectory.
OpenVSLAM is different when compared with other vSLAMs; this is because it can treat omnidirectional videos.
Omnidirectional images contain information in all directions, and structural patterns such as those of buildings that usually appear on the left and right side of the camera which is directed to move along the streets.
Therefore, OpenVSLAM can work with a high number of visual features that exist in an omnidirectional video for accurate estimation.

We performed vSLAM on each video independently.
The camera locations derived by OpenVSLAM from the omnidirectional videos were determined to be reasonably accurate.
We used two reference points for each video and aligned the corresponding camera locations with the map coordinates.

Generally, vSLAM accumulates errors, which result in a scale drift problem~\cite{scale_drift}.
The vSLAM technique addresses this problem via loop closing~\cite{loop_closing}, which is a constraint that the camera imposes, and in which localized visual features coincide at the loop-closing.
However, each video in our study is captured along a street and does not possess any loop closures.
A loop closing algorithm is therefore not applicable to our videos.

\subsection{Photometric Reconstruction}

These days, several images or videos about a specific area can be captured by using mobile devices~\cite{pedestrian_navigation}.
They contain a lot of information, while the original data collection is difficult to understand because it is not structured~\cite{world_photo}.

In previous works, analysis of the spatial relationship among data, and intuitive transitions between sources enable users to experience a virtual tour in the target area~\cite{photo_tourism}~\cite{video_scape}~\cite{street_slide}~\cite{image_based_exploration}.
Photo tourism~\cite{photo_tourism} calculates the relative camera poses of collected images by using structure from motion (SfM)~\cite{sfm}.
On the interface, it visualizes the spatial relationship and switches the images by user input.

In our study, the spatial relationship between two source videos is represented as one intersection.
By limiting the points of the video switching at intersections, the relationships are represented in a simple manner.

\begin{figure}[t]
\begin{center}
  \includegraphics[keepaspectratio,scale=0.3]{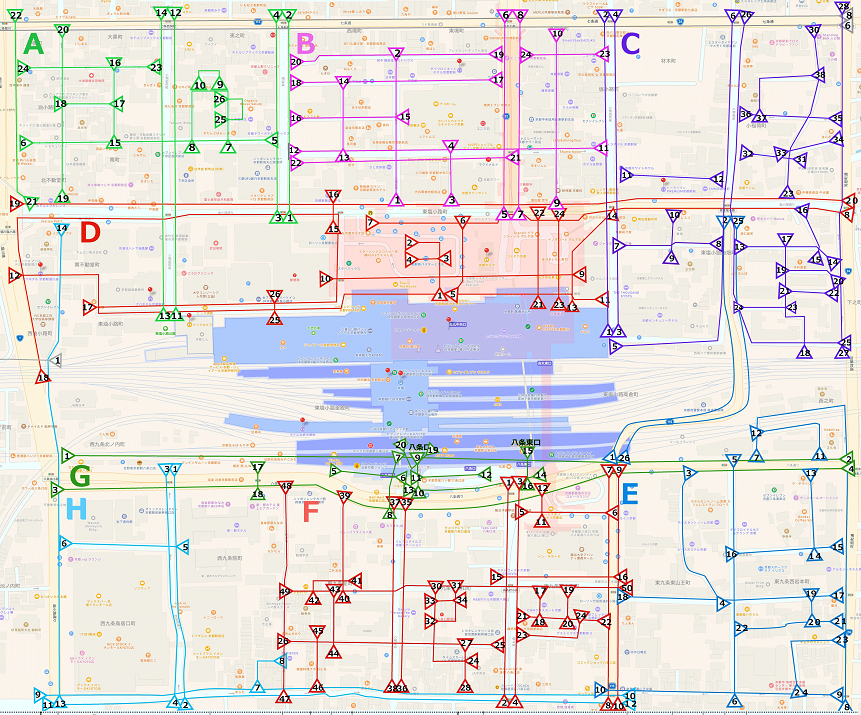}
  \caption{One of the shooting areas ($1km^2$ around Kyoto station). The lines show the streets we captured.}
\label{capture_lines}
\end{center}
\end{figure}

\begin{figure*}[t]
\begin{center}
  \includegraphics[keepaspectratio,scale=0.40]{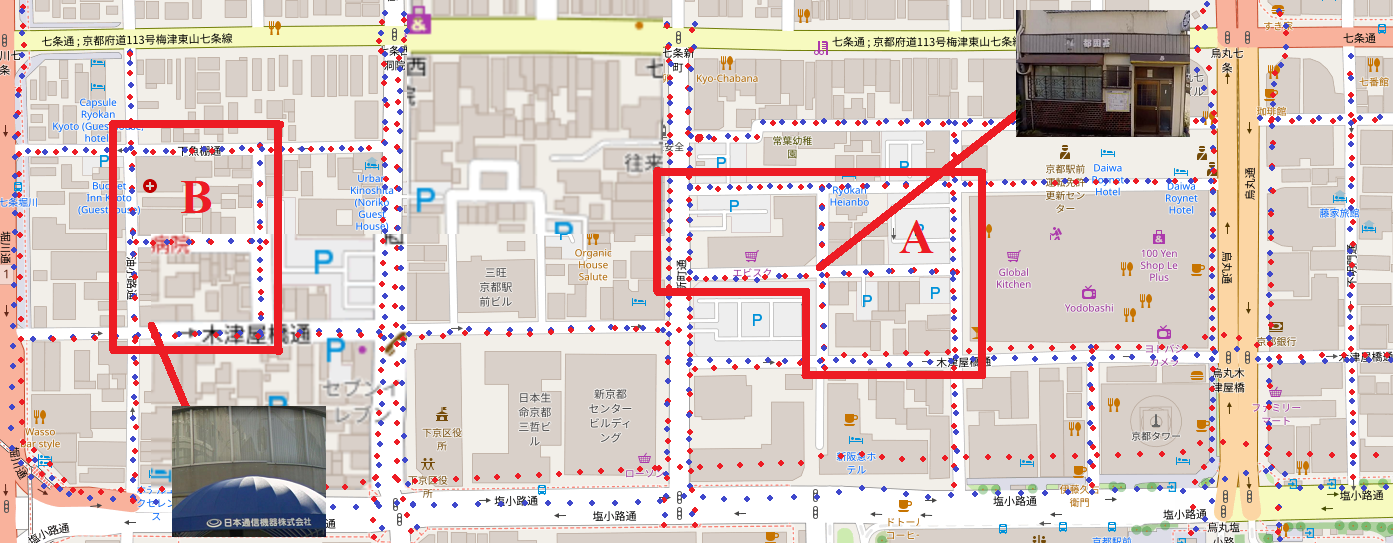}
  \caption{Mapping of trajectories onto a common map. The areas marked A and B indicate the search area in the exploring experiment described in Section 5.2.}
\label{mapping_example}
\end{center}
\end{figure*}

\section{Movie Map Building System}

In this section, we describe the processing flow of video data.
Figure \ref{system_abstract} shows the outline of our Movie Map system.
The map building system is divided into three stages: acquisition, analysis, and management of data.
These are further subdivided into several processes.
To build a Movie Map in practice, we only have to input the captured omnidirecional videos and two reference coordinates per video into the system, and the system automatically outputs the materials that are used in our exploring interface.
Our system is easily applicable to various areas.

\subsection{Data Acquisition}

We captured omnidirectional videos along the streets in the target area, and then manually assigned coordinate reference information.

\subsubsection{Acquisition of Street Videos}\,

\noindent
We captured street videos using an omnidirectional camera. This was achieved by a person carrying the omnidirectional camera and walking on the streets.
We captured videos along streets in areas surrounding the Kyoto station and Namba station.
The overall picture of the area with a size of $1km^2$ around Kyoto station is depicted in Figure. \ref{capture_lines}.
We physically walked in both directions to capture videos of a single street. 
Acquiring two-way-videos is necessary for our task of producing a natural feel when moving forward and backward on the map.
Moreover, it should be noted that the streets did not need to be straight; thet could be curved or turning streets. 
However, we assumed that two shooting paths intersect at one point.
If they intersect at multiple points, the subsequent analysis fails.
In this case, we divided one of them into two paths to prevent our analysis from failing.

\subsubsection{Assignment of Reference Coordinates}\,

\noindent
We assigned the reference point coordinates to each captured omnidirectional video.
As described in Section 3.2.1, we applied vSLAM to each video and estimated the relative coordinates of the camera positions in the frames of each video. 
In order to integrate all camera positions into a single map, we assigned global coordinates that were common to all captured videos.  

We assigned information on the latitude and longitude of two reference points, which corresponded to the start and end frame of each video.
Instead of latitude and longitude, we could have also used specific global coordinates defined on a map, as long as they were the same for all captured videos.

\subsection{Data Analysis}

In order to create a route, the intersections of the streets were considered to be the most important points.
A route map was represented by the information of its streets and their intersections.
In this section, we analyze the captured street videos, and automatically obtain their intersection information. 

\subsubsection{Application of vSLAM}\,

\noindent
We estimated the relative camera poses, including positions and orientations, using OpenVSLAM.
The accuracy of the estimations from OpenVSLAM was high as the program used visual features in all directions in the omnidirectional image to optimize the actual camera positions.
The error is practically negligible for general path lengths, such as those shown in Figure. \ref{capture_lines}.

\subsubsection{Mapping Videos onto Common Coordinate}\,

\noindent
We mapped all relative camera positions onto a common coordinate space.
Camera positions, as estimated by OpenVSLAM, were independently calculated for each video.
We mapped all videos onto a common coordinate system to associate them with each other.

In that time, we use the reference point coordinate information mentioned in Section 3.1.2.
By considering the vector from the start point to the endpoint of the estimated camera positions, we obtained the rotation and scaling so that the vector could be equal to a vector between the reference start point and end point coordinates.
We also applied the same rotation and scaling to all camera positions and directions in a video.
Furthermore, we translated all camera positions so that the start point coordinate matched the reference start point coordinate.
Repeating this process for all videos, we aligned all camera positions on a common coordinate of the map.

Figure \ref{mapping_example} presents an example of this type of mapping.
Different color markers represent the coordinates of key-frames in both ways of one street.

\subsubsection{Intersection Detection}\,

\begin{figure}[t]
  \begin{tabular}{cc}
    \begin{minipage}{0.5\columnwidth}
      \begin{center}
        \includegraphics[scale=0.15]{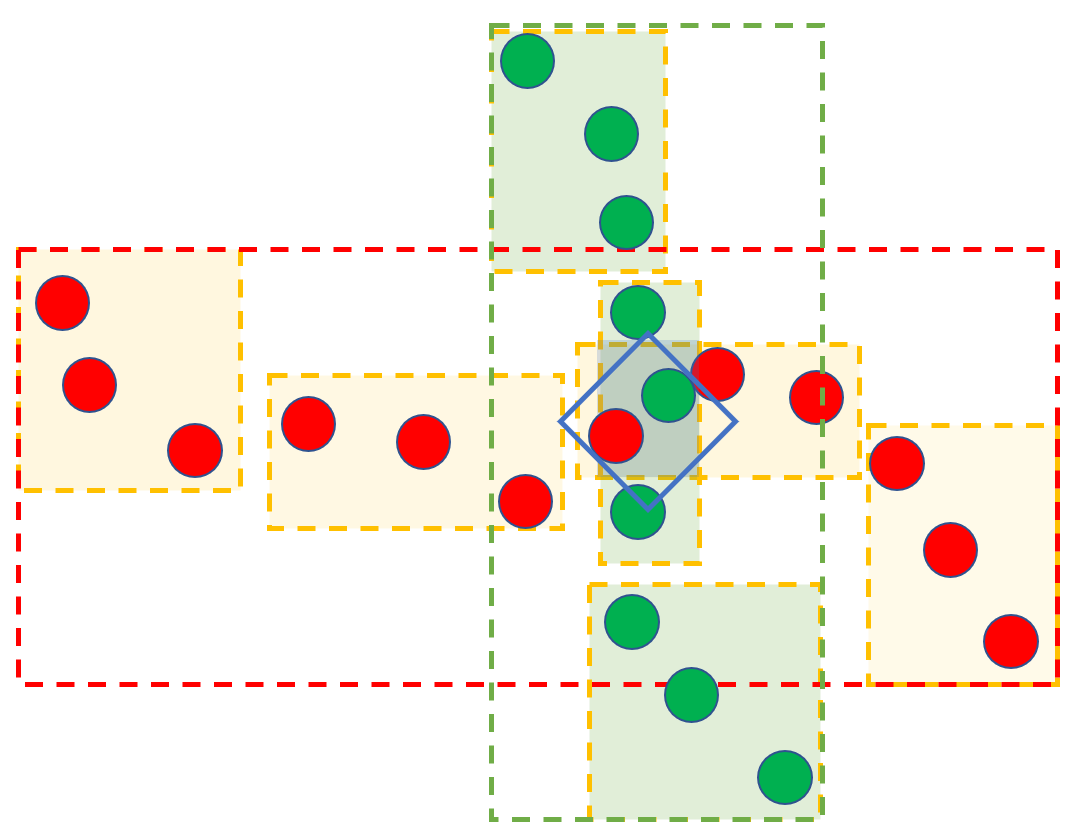}
        \subcaption{Detection Procedure}
        \label{detection_procedure}
      \end{center}
    \end{minipage}
    \begin{minipage}{0.5\columnwidth}
      \begin{center}
        \includegraphics[scale=0.15]{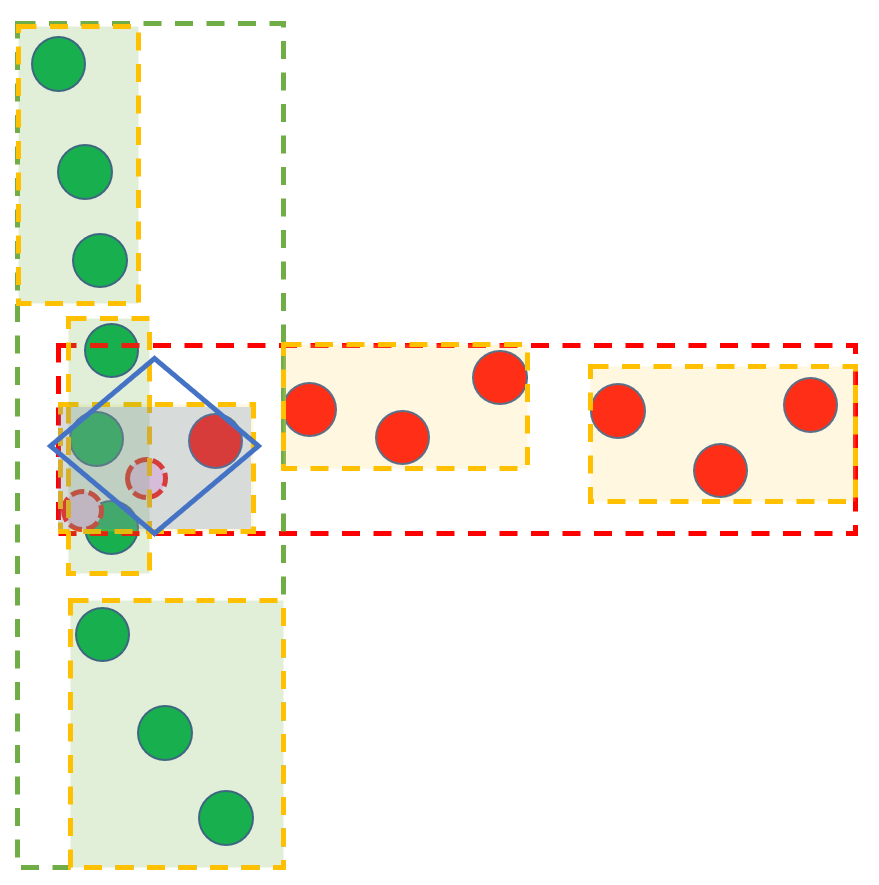}
        \subcaption{Extended rectangle for the edge of the route}
        \label{t_junction}
      \end{center}
    \end{minipage}
  \end{tabular}
  \caption{Intersection detection}
\label{intersection_detection}
\end{figure}

\noindent
We detect the intersection information using the coordinate information mapped onto the common map, and the visual features.
The intersection information here refers to which street video intersects with which video in which frame, coordinates of the intersection, and the relative rotation between the frames of these two videos.
We first find video pairs that intersect each other.
Suppose video A and video B have an intersection, then we can determine the frames of each video that are most similar in location and visual features.

For a quick detection, each trajectory was divided into rectangles and the intersection candidate frames were narrowed down based on their overlap.
The procedure of this method is shown in Figure. \ref{intersection_detection}.
As shown in Figure \ref{detection_procedure}, two trajectories captured in different street videos are divided every hundred frames.
Next, rectangles that covered the divided parts were taken into further consideration and were then searched for overlapping rectangular pairs between the trajectories.
Finally,. we found the frames with the least distance between two trajectories by implementing a full search inside the overlapping rectangle.

If the end of the route forms an intersection, as shown in Figure. \ref{t_junction}, the detection may not be successful due to a lack of rectangular overlap, depending on the start points and endpoints. 
In this case, as an exception process at the time of splitting, the start points and endpoints were extended by several hundred frames, to which a rectangle was added that covered the extended part.

As this search was performed by using only the location information obtained by mapping the camera positions, which was estimated by vSLAM based on the reference points, it contained a few errors.
To minimize these errors, we adjusted the intersection frames using their visual feature similarity.
For dozens of frames around the location, which was based on the detected intersection frame, we rotate the frames in the same direction based on the camera position estimated by vSLAM.
Thereafter, we extracted the ORB features~\cite{orb} for each image and determined the frame pairs with the highest similarity. 
These pairs were set as the correct intersection frames.
When using visual similarity, the results are more accurate than those obtained when only using location information.

For the intersection frames obtained as a result of detection and adjustment, we recorded the following data: the two videos to which the frames belong to, the timestamp of the frames in the videos, coordinates, and the relative camera rotation between both frames. 

We automatically repeated this detection for all pairs of streets and obtained information from all intersections in the target area.
\subsection{Data Management}

\begin{figure}[t]
\begin{center}
  \includegraphics[keepaspectratio,scale=0.22]{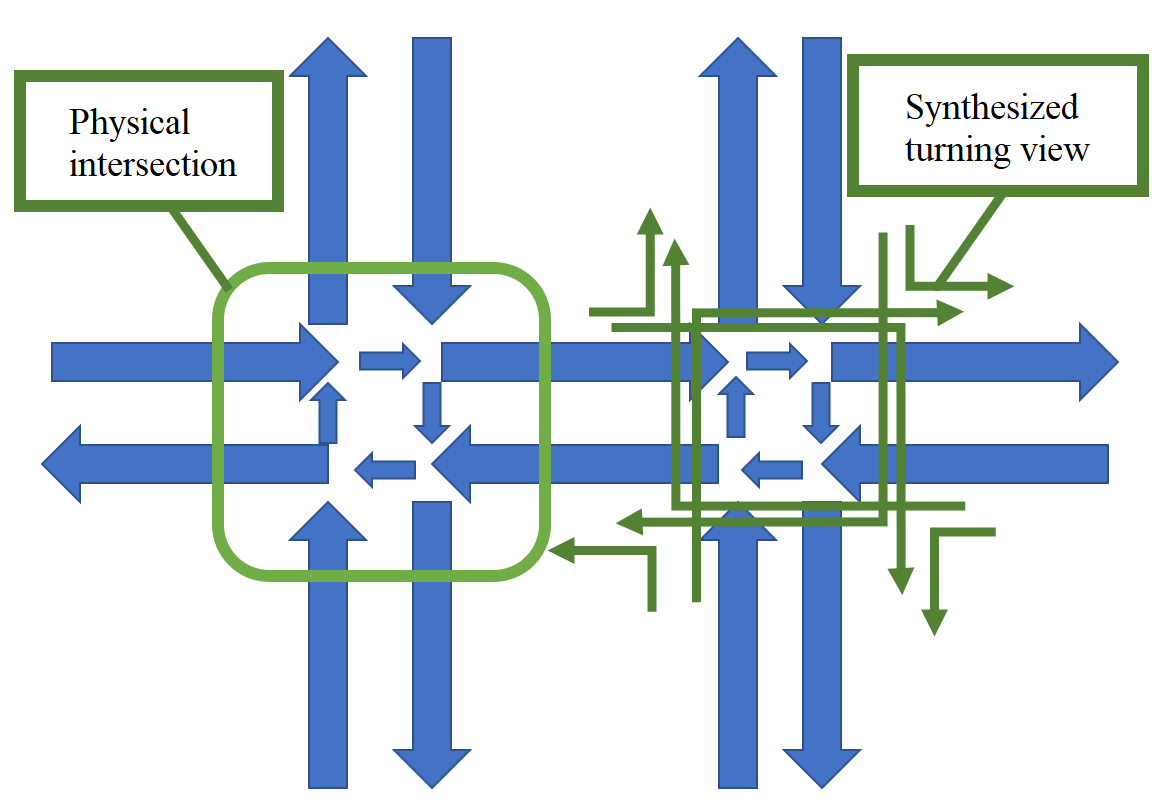}
  \caption{Split and synthesis at a physical intersection}
\label{management}
\end{center}
\end{figure}

Using the intersection information obtained in the data analysis stage, it was possible to synthesize a route movie by editing and playing back a part of the videos.
In order to produce an interactive real-time player, we converted the videos into a format used on the interface.
According to the intersections of the streets, we segmented the street videos into sections between intersection frames and added metadata to specify the video sections.
In advance, we synthesized the turning views at intersections, which were inserted to produce a natural transition from one video section to another.

\subsubsection{Splitting Street Videos into Sections}\,

\noindent
We split the street videos at intersections on the path.
When we used the video in our interface, the unit of playing is a section between intersections.
It was significantly faster to load a section than to load the full street video and specify which part was to be played after.
The division was performed according to the analyzed intersection frame information, which included the street video index and timestamp.

There are several sections corresponding to a physical intersection because we maintain videos of both directions.
As shown in Figure. \ref{management}, they intersect complicatedly and small sections are segmented in the intersection.
 
For each section data, we added the following metadata: the street video ID to which it belongs and the intersection IDs of both ends of the section.

\subsubsection{Synthesis of Turning Views at Intersections}\,

\noindent
We synthesized the turning views for each intersection as shown in Figure. \ref{interface_example}.
In our interface, the user can turn at any intersection.
Before and after turning, we switched from one video section to another video section.
However, switching movies without any interpolation made us feel uncomfortable and could cause inconsistencies in our cognition of position and direction, as presented in Section 5.1.

Lippman~\cite{movie_map} captured different turning movies for each intersection and inserted them whenever the driver turned.
However, we needed eight turning patterns at a standard physical intersection as depicted in Figure. \ref{management}.
When we considered the number of intersections that existed in a certain area, shooting videos for all of them were deemed to be not feasible.
Owing to the use of omnidirectional videos, we could easily synthesize the turning views by rotation. 

Video sections before and after the intersection were from different street videos captured at different times.
Hence, it can be often observed that the brightness and objects, such as cars and people before and after the intersection change. 

Therefore, we synthesized the turning views by blending the intersection frames; the frame before turning is denoted as frame $I$ and the frame after turning is denoted as frame $J$.

As the camera directions of the frames are mapped onto a common map, as described in Section 3.2.2, we can use the rotation angles between frames $I$ and $J$.
In our generating method, we rotated frame $J$ to align it with frame $I$ and then proceeded to blend both frames.
We synthesized a turning view by rotating the blended frames; consequently, the blending ratio linearly changed from 0 to 1.
As a result, we could synthesize a turning view starting from frame $I$ and gradually change it to frame $J$ during the rotation.

\section{Movie Map Exploring Interface}

\begin{figure}[t]
  \begin{tabular}{cc}
    \begin{minipage}{0.5\columnwidth}
      \begin{center}
        \includegraphics[scale=0.22]{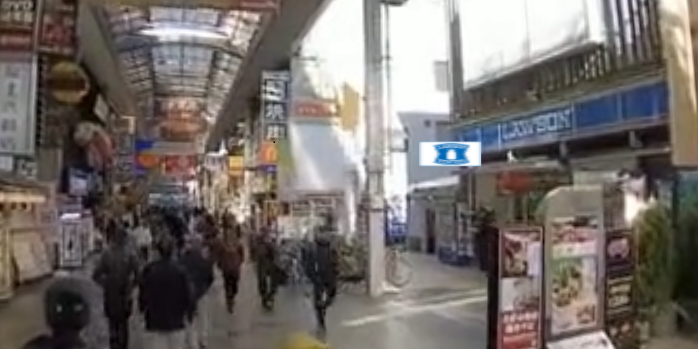}
        \subcaption{Distant billboard}
      \end{center}
    \end{minipage}
    \begin{minipage}{0.5\columnwidth}
      \begin{center}
        \includegraphics[scale=0.22]{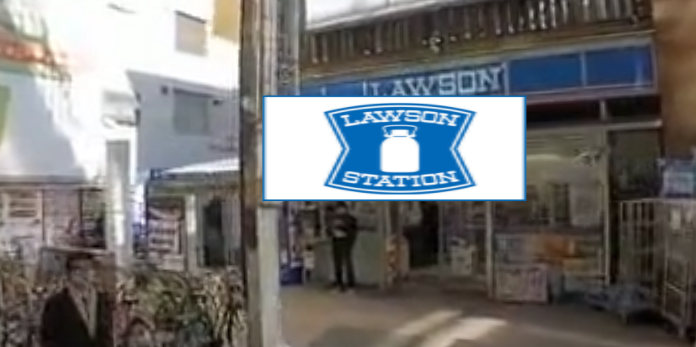}
        \subcaption{Close billboard}
      \end{center}
    \end{minipage}
  \end{tabular}
  \caption{Example of virtual billboard}
\label{virtual_billboard}
\end{figure}

\begin{figure}[t]
\begin{center}
  \includegraphics[keepaspectratio,scale=0.4]{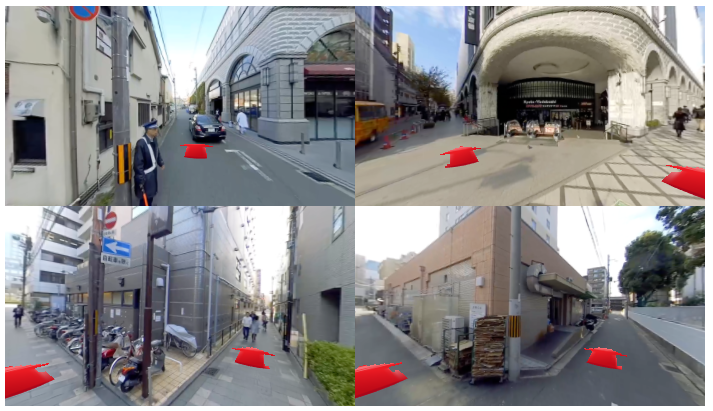}
  \caption{Select start point to explore}
\label{interface_start}
\end{center}
\end{figure}

\begin{figure}[t]
\begin{center}
  \includegraphics[keepaspectratio,scale=0.28]{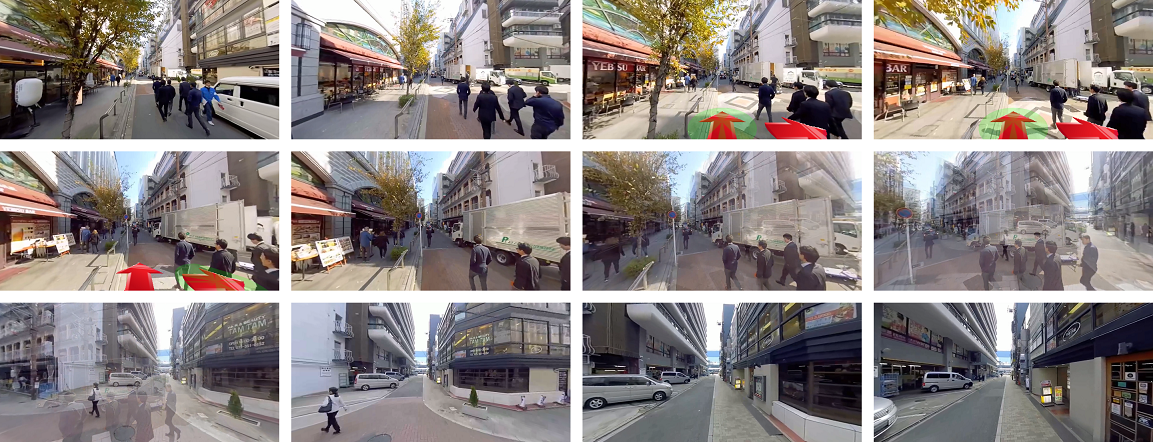}
  \caption{An example of a sequence of a synthesized route movie}
\label{interface_example}
\end{center}
\end{figure}

\begin{figure*}[t]
 \begin{minipage}{0.16\hsize}
  \begin{center}
   \includegraphics[keepaspectratio,scale=0.22]{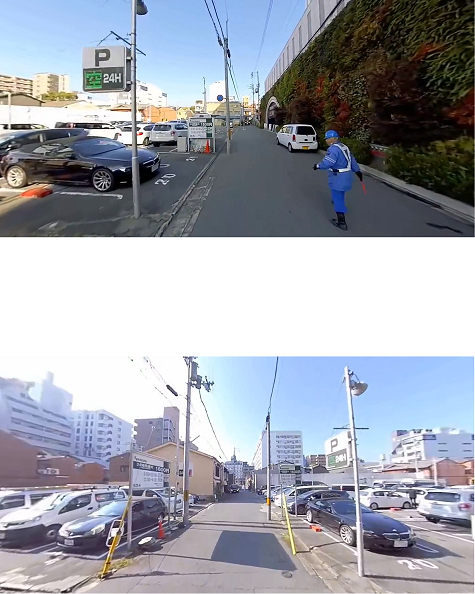}
   
   (a) Intersection 1 Method A
  \end{center}
 \end{minipage}
  \begin{minipage}{0.16\hsize}
  \begin{center}
   \includegraphics[keepaspectratio,scale=0.22]{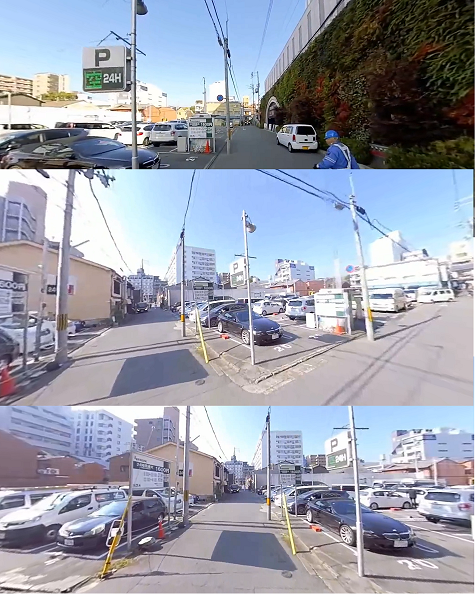}
   
   (b) Intersection 1 Method B
  \end{center}
 \end{minipage}
 \begin{minipage}{0.16\hsize}
  \begin{center}
   \includegraphics[keepaspectratio,scale=0.22]{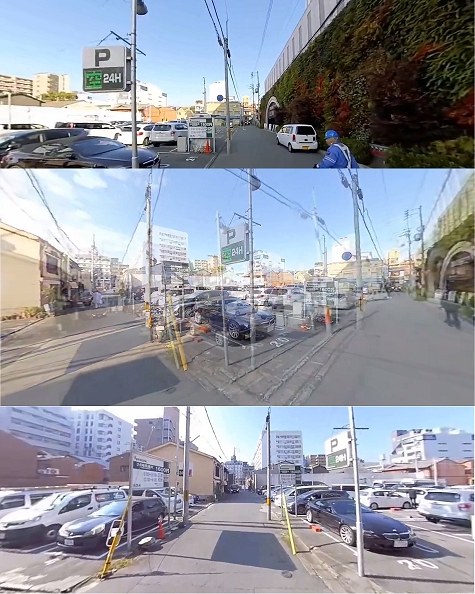}
   
   (c) Intersection 1 Method C
  \end{center}
 \end{minipage}
  \begin{minipage}{0.16\hsize}
  \begin{center}
   \includegraphics[keepaspectratio,scale=0.22]{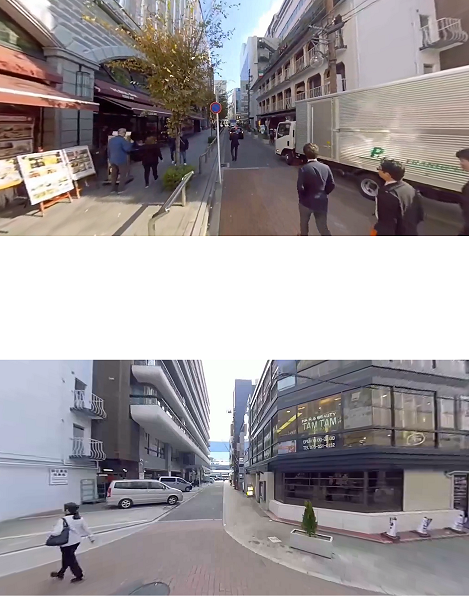}
   
   (d) Intersection 2 Method A
  \end{center}
 \end{minipage}
  \begin{minipage}{0.16\hsize}
  \begin{center}
   \includegraphics[keepaspectratio,scale=0.22]{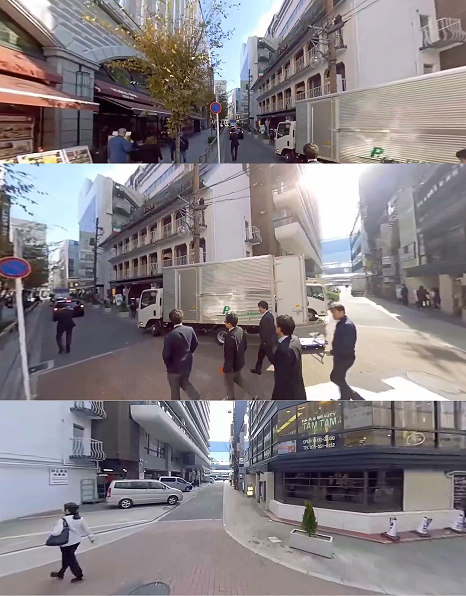}
   
   (e) Intersection 2 Method B
  \end{center}
 \end{minipage}
 \begin{minipage}{0.16\hsize}
  \begin{center}
   \includegraphics[keepaspectratio,scale=0.22]{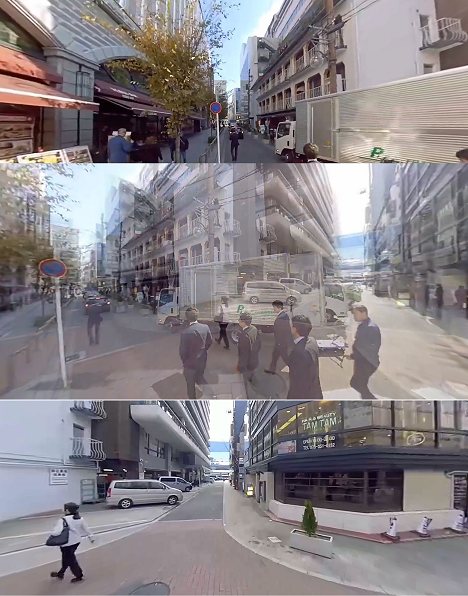}
   
   (f) Intersection 2 Method C
  \end{center}
 \end{minipage}
 \caption{Turning views in the experiment}
 \label{inters}
\end{figure*}

We propose an exploration-type interface based on the Movie Map made of omnidirectional street videos that are analyzed and managed as described in the previous sections.
This interface is useful for tasks such as prior learning and conducting a virtual tour of a target area.
Figure \ref{interface} shows an overall view of the interface.
On this interface, the user can display street videos that correspond to the coordinates on the background map.
By selecting the desired direction at an intersection, we can easily manipulate the directions of each movement. 
An example of the route switching movie at an intersection is presented as a slide show in Figure. \ref{interface_example}.

\subsection{Interaction on the Interface}

Using the metadata of the intersections, the interface draws intersection points on the map shown in the background.
Figure \ref{interface_start} is a screen for determining the starting point of this exploration.
On this screen, landmarks pre-defined in the target area, and an arrow indicating the direction of walking are displayed.
The user selects one of the landmarks and a walking direction to start the exploration.
When the user starts exploring, a screen for playing the walking video appears, and the street video starts to play.
The user can change the viewing direction by dragging the omnidirectional video and can easily specify the walking direction at the intersections by choosing the arrow bottoms that appear when approaching an intersection.
By clicking on one of the arrows, the user can easily select the desired direction and the view is turned accordingly and switched into the next video section.
Moreover, the user can perform basic video operations such as a change of playback speed by interacting with the button inputs.
In this way, the user can walk continuously in a chosen area at his or her desired speed.

\subsection{Hiding Details of Intersections for Visualization}

In the interface shown in Figure. \ref{interface}, the intersection on the route map is visualized as a single point. The actual data, as described in Section 3.1.1, consists of two-way videos of a street; one physical intersection generally contains four intersection frame pairs. 
Multiple intersection frame pairs of a single physical intersection are grouped as a point, and its cluster center is drawn as the location of the point on the map.

We keep the connections of the grouped intersections as a graph structure.
Based on this data, when approaching an intersection, the system displays the navigation arrows only for the directions in which the physical streets exist.
Using the location of the frames, the system visualizes the current position on the map.

\subsection{Virtual Billboards}

We can display virtual billboards in the view of the Movie Map.
After building the Movie Map, the video is aligned in the direction of the street.
We can overlay a billboard in the video by corresponding it to its specific location, as shown in Figure. \ref{virtual_billboard}.
The billboard can be shown from near to far viewpoints.
To show the billboard, we need to manually specify a position.
A single location per billboard is kept in a billboard list.
We use the time stamp of the video for the location.
When the user approaches the location, the corresponding billboard pops up.
Each billboard has additional information, which appears by clicking the virtual billboards.

\section{Experiment -- User Studies}

\begin{table*}[htb]
  \includegraphics[keepaspectratio,scale=0.28]{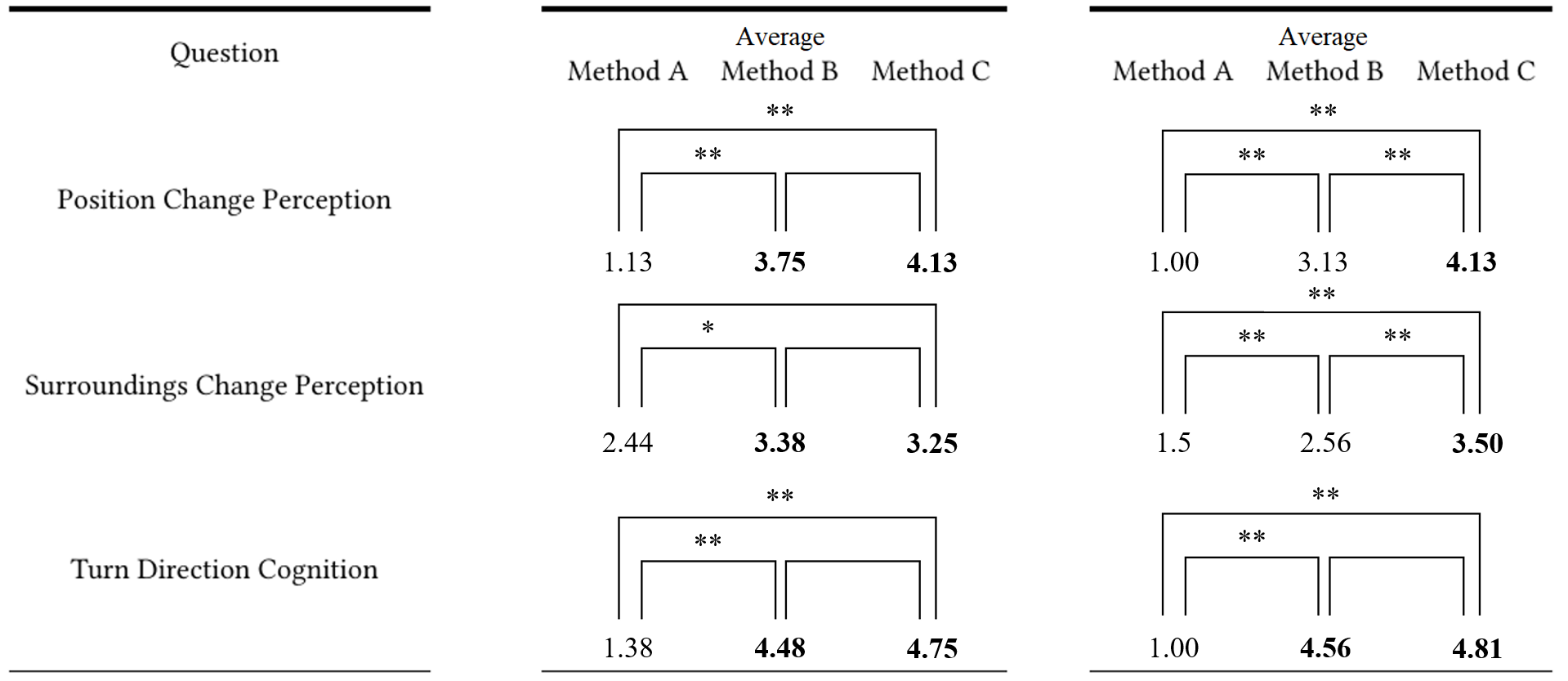}
  
\hspace{180pt} (a) Intersection 1 \hspace{70pt} (b) Intersection 2 \hspace{40pt}
  
  $* : p < 0.05, ** : p < 0.01$
  
  \vspace{0.5pt}
  \caption{Evaluations of three methods of turning view synthesis}
  \label{result1}
\end{table*}

\begin{table}[htb]
   \includegraphics[keepaspectratio,scale=0.17]{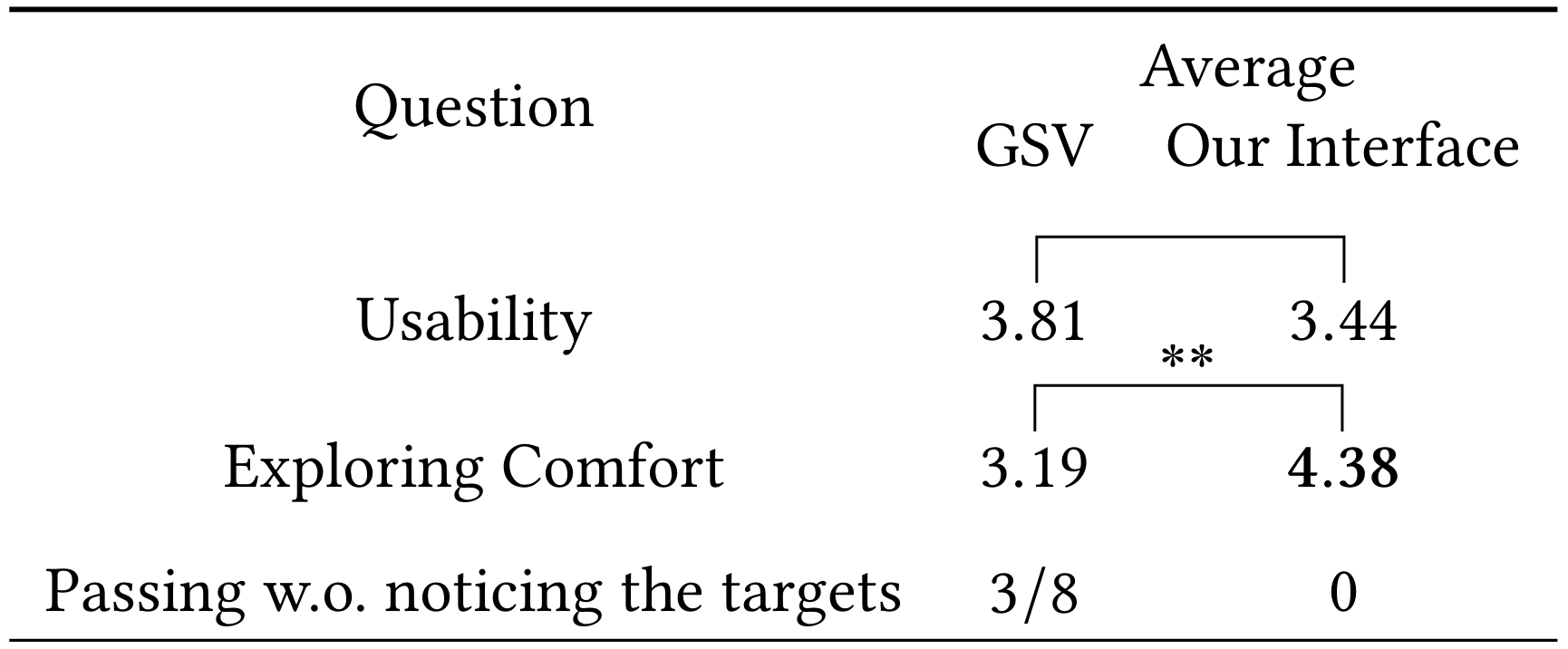}
  
  $** : p < 0.01$
  
  \vspace{0.5pt}
  \caption{Results of comparisons between GSV and our system}
  \label{result2}
\end{table}

We produced two user studies for the evaluation of our Movie Map. 
(1) We evaluated the intersection movies, i.e., the degree of feeling visual discontinuity when changing routes at intersections. (2) We evaluated the usability of the proposed interface and subsequently compared it to GSV. In both cases, the ratings were based on user studies.
We did not include virtual billboards in the interface in the user studies.
The subjects were 16 students who had no prior knowledge of the study.

\subsection{Evaluation of Synthesis of Intersection Movie}

We evaluated the effect of the insertion of the synthesized turning view on the visual discontinuity.

Subjects watched three different videos of route changes at two intersections, as shown in Figure 10: method A--the two street videos were switched directly without any processing; method B--one of the intersection frames was inserted with a rotation, and method C,--the blended turning view described in this paper was inserted.

Subjects answered questions regarding the perceived naturalness of the change of position and the surroundings, and the ease of cognition in turning directions.
The score consisted of a 5-point scale; positive: 5, weakly positive: 4, neutral: 3, weakly negative: 2, negative: 1.

The results are shown in Table \ref{result1}.
We used two intersections for comparisons.
Intersection 2 was more difficult case than intersection 1.
Intersection frames at intersection 1 were similar in terms of their position and surroundings.
However, in intersection 2, the videos before and after the switch were captured at different times during the day.
Furthermore, the locations of the intersection frames were not close enough and the surroundings changed; thus, the appearance of the intersection frame pairs was not similar. 

From these results, method A was considered unacceptable at both intersections.
Comparing methods C and B, method C performed significantly better than method B at intersection 2 in terms of the three evaluation criteria.
While, at intersection 1, methods C and B did not have a significant difference.

At intersection 1, which is a simpler intersection, the rotation was considered to be enough to produce views for natural transitions and cognition of turning direction.
However, at intersection 2, which is harder because of large changes in the position and surrounding appearance before and after the transition, method C was evaluated to be significantly better than B because of the blending rotation turning, which made the transition smoother.

\subsection{Evaluation of the Interface Comparing with GSV}

We compared our system with GSV under a scenario wherein subjects explored a small area and searched for a target in that area using the interface.

We prepared two areas and targets---area A and area B, both with targets, as represented in Figure. \ref{mapping_example}.
The subjects explored A or B by using Movie Map or GSV.
For example, subject 1 explored area A using Movie Map and area B using GSV to avoid any learning effects. 

They evaluated both interfaces in terms of their usability for exploring and the degree of comfort when exploring.
A scale of five points was used for their evaluation, whereby five is the best and one is the worst.

The results of this evaluation are shown in Table \ref{result2}.
As for the usability, there were no significant differences between GSV and our interface.
It may be due to experiences of the subjects.
Most of them had prior experience with GSV and were not familiar with our interface, which required a real-time input of the directions for walking.
Although the usability did not change, no one overlooked the target by our interface in area A.
However, three of the eight subjects who explored area A by GSV overlooked the target at least once.
In fact, one of the subjects could not find the target in the limited exploration time of 3 minutes.

Nonetheless, the proposed interface was rated higher owing to its feeling of exploration.
This was because playing continuous videos and using natural transitions at intersections resulted in a more natural feeling akin to walking.

\section{Conclusion}

We proposed a new Movie Map system and its exploration interface.
The proposed method involved the acquisition, analysis, management, and interaction of omnidirectional videos.
Once the street videos were acquired, the entire pipeline of processing was almost automated completely.

The proposed system segmented videos into sections using the information of the detected intersections.
Moreover, intersection turning views were synthesized in advance.
The walking videos of the target area were displayed by playing these videos according to the specified route.
The manipulation of the interface consisted of a simple selection of the directions at intersections.
Moreover, with additional operations, we could display virtual billboards.

In the experiments, we compared three types of intersection movies, which included the generated turning views and the use of a blended turning view, which was shown to be effective to some extent.
Additionally, we compared our interface with GSV and determined that the proposed interface provided the user with a better experience when exploring.


\bibliographystyle{ACM-Reference-Format}
\bibliography{all}

\end{document}